\documentclass[11pt]{article}

\usepackage[usenames,dvipsnames,table]{xcolor}

\usepackage{changepage}

\usepackage{jheppub} 

\usepackage{graphicx}
\usepackage{amsmath, amssymb}

\usepackage{manfnt}

\newcommand{\be}{\begin{eqnarray}}
\newcommand{\ee}{\end{eqnarray}}
\newcommand{\bea}{\begin{eqnarray}}
\newcommand{\eea}{\end{eqnarray}}

\newcommand{\bn}{\begin{enumerate}}
\newcommand{\en}{\end{enumerate}}

\def\e{\epsilon}

\def\Z{{\mathbb Z}}
\def\C{{\mathbb C}}

\title{${\cal N}=1$ conformal duals of  gauged $E_n$ MN models}

\author[a]{Shlomo S. Razamat}
\author[b]{and Gabi Zafrir}

\affiliation[a]{Department of Physics, Technion, Haifa, 32000, Israel}
\affiliation[b]{Dipartimento di Fisica, Universita di Milano-Bicocca \& INFN, Sezione di Milano-Bicocca,
I-20126 Milano, Italy}

\emailAdd{razamat@physics.technion.ac.il}
\emailAdd{gabi.zafrir@unimib.it}

\abstract{We suggest three new ${\cal N}=1$ conformal dual pairs. First, we argue that the ${\cal N}=2$ $E_6$ Minahan-Nemeschansky (MN) theory with a $USp(4)$ subgroup of the $E_6$ global symmetry conformally gauged with an ${\cal N}=1$ vector multiplet and certain additional chiral multiplet matter resides at some cusp of the conformal manifold of  an $SU(2)^5$ quiver gauge theory. Second, we argue that the  ${\cal N}=2$ $E_7$ MN theory with an $SU(2)$ subgroup of the $E_7$ global symmetry conformally gauged with an ${\cal N}=1$ vector multiplet and certain additional chiral multiplet matter resides at some cusp of the conformal manifold of a conformal ${\cal N}=1$ $USp(4)$ gauge theory. Finally, we claim that the ${\cal N}=2$ $E_8$ MN theory with a $USp(4)$ subgroup of the $E_8$ global symmetry conformally gauged with an ${\cal N}=1$ vector multiplet and certain additional chiral multiplet matter resides at some cusp of the conformal manifold of an ${\cal N}=1$ $Spin(7)$ conformal gauge theory. We argue for the dualities using a variety of non-perturbative techniques including anomaly and index computations. The dualities can be viewed as ${\cal N}=1$ analogues of ${\cal N}=2$ Argyres-Seiberg/Argyres-Wittig duals of the $E_n$ MN models. We also briefly comment on an ${\cal N}=1$ version of the Schur limit of the superconformal index. 
}

\begin{document} 

\maketitle
\flushbottom

\section{Introduction} 

Strongly coupled supersymmetric conformal field theories (SCFTs) can be engineered in a variety of ways. In particular they can be obtained as descriptions of an infra-red (IR)  fixed points of renormalization group (RG) flows starting from a small relevant deformation of a weakly-coupled SCFT. The canonical examples in $4d$ are the flows starting with  ${\cal N}=1$ SQCD in the conformal window. Strongly coupled SCFTs can be also engineered starting from a weakly coupled SCFT with a  conformal manifold and tuning the couplings to be large. Canonical examples here include the ${\cal N}=4$ SYM and a variety of ${\cal N}=2$ conformal gauge theories. Moreover, a weakly coupled SCFT can have strongly-coupled loci, cusps, on the conformal manifold which can be alternatively described  by weak gauging of a global symmetry of some strongly-coupled SCFT. A paradigmatic example of this is given by the ${\cal N}=2$ Argyres-Seiberg duality \cite{Argyres:2007ws}. In fact the discovery of these dualities triggered,  starting with \cite{Gaiotto:2009we}, an avalanche of new understandings of the dynamics of strongly-coupled ${\cal N}=2$ SCFTs.

Conformal manifolds with minimal supersymmetry, ${\cal N}=1$ as opposed to ${\cal N}\geq 2$, in four dimensions have been much less studied. However, the existence of interesting conformal field theories with a manifold of exactly marginal couplings was established quite some time ago \cite{Leigh:1995ep} (see {\it e.g.} \cite{Sohnius:1981sn,Howe:1983wj,Parkes:1984dh} for earlier works), and the technology to identify such models is rather straightforward \cite{Green:2010da} (see also   \cite{Kol:2002zt}). 
One of the interesting features accompanying conformal manifolds with extended supersymmetry is that different regions of it might be describable by different looking weakly coupled, or partially weakly coupled, models as already mentioned above. In fact in \cite{Razamat:2019vfd} numerous such dualities even for ${\cal N}=1$ cases were suggested.  The algorithm to search for such dual pairs used in \cite{Razamat:2019vfd} is rather simple\footnote{This algorithm can be thought of as ${\cal N}=1$ generalization of the search for ${\cal N}=2$ dualities explored by Argyres and Wittig \cite{Argyres:2007tq}.}. Assuming the dual descriptions of a given model is conformal, that is  no RG flow is involved, significantly restricts the space of possibilities. In particular, if one seeks for a conformal gauge theory description, the two conformal anomalies, $a$ and $c$, completely fix the dimension of the gauge group and the dimension of the representation of the matter fields. This leaves only a finite set of possibilities to go over in the search for a dual description, which surprisingly often actually results in finding such a putative dual.

For the algorithm above to be applicable the model at hand should possess an ${\cal N}=1$ preserving conformal manifold. However, many interesting SCFTs do not have such manifolds. Maybe some of the most well known examples are the ${\cal N}=2$ Minahan-Nemeschansky (MN) $E_n$ SCFTs \cite{Minahan:1996fg,Minahan:1996cj}\footnote{In recent years however some of these models have been constructed  starting with weakly coupled gauge theories using RG flows \cite{Gadde:2010te,Gadde:2015xta,Agarwal:2018ejn,Zafrir:2019hps}.}. However, one can use such SCFTs as components of larger models with a conformal manifold. A way to do so is to couple the conserved currents of a subgroup of the global symmetry to dynamical vector fields and add sufficient amount of matter so that the gauging, as well as any needed superpotential interactions, will be exactly marginal. One can do so for example for $E_n$ MN models  preserving the ${\cal N}=2$ supersymmetry \cite{Argyres:2007ws,Argyres:2007tq}. Once a conformal manifold appears, alongside comes the possibility that somewhere on it a dual weakly coupled description emerges. This was indeed the case for ${\cal N}=2$ gaugings of MN models discussed in \cite{Argyres:2007ws,Argyres:2007tq}. 

In the current note we will start from $E_{6,7,8}$ MN model and construct theories with conformal manifolds by gauging subgroups of the global symmetry, as in \cite{Argyres:2007ws,Argyres:2007tq}, but now preserving only ${\cal N}=1$ supersymmetry.  We will argue that after such a gauging somewhere on the conformal manifold a dual ${\cal N}=1$ conformal gauge theory description emerges. For the $E_6$ case we will find a dual description as an $SU(2)^5$ quiver gauge theory while in the $E_{7,8}$ cases the dual will be a gauge theory with a simple gauge group\footnote{For other interesting dualities between ${\cal N}=1$ gauge theories and constructions involving more general class ${\cal S}$ models \cite{Gaiotto:2009we} see for example \cite{Gadde:2013fma}.}.
In each case we will test the dualities by studying properties which are invariants of the conformal manifold, like anomalies and superconformal indices.
We suspect that there should be a powerful geometric interpretation  (constructing the models starting  from $6d$ SCFTs on Riemann surfaces, {\it e.g.}  for some examples see \cite{Gaiotto:2009we,Benini:2009mz,Bah:2012dg,Gaiotto:2015usa,Razamat:2016dpl,Kim:2017toz,Razamat:2018gro,Pasquetti:2019hxf,Razamat:2019ukg}, or utilizing other string theory constructions) of the results presented here, as well as the ones reported in \cite{Razamat:2019vfd}. We leave this aspect for future investigations.

 \section{Dual of $E_6$  MN theory with $USp(4)$ subgroup gauged}
 
 Let us consider  the Minahan-Nemeschansky $E_6$ SCFT \cite{Minahan:1996fg}. We consider the branching of representations of the $E_6$ symmetry to representations of its $U(1)_a\times SO(10)$ maximal subgroup such that ${\bf 27}\to {\bf 1}_{-4}\oplus{\bf 10}_{2}\oplus {\bf 16}_{-1}$, and further decompose  $SO(10)$ to $USp(4)_g\times USp(4)$ such that
 \be
 &&{\bf 10}\to ({\bf 5},{\bf 1})\oplus ({\bf 1},{\bf 5})\,,\qquad 	{\bf 16}\to ({\bf 4},{\bf 4})\,.
 \ee  Then we gauge the $USp(4)_g$ symmetry with the addition of six  fundamentals, $q_L$, and three two index traceless antisymmetrics, $\phi_A$.  Note that the imbedding indices of $SO(10)$ in $E_6$ and of $USp(4)_g$ in $SO(10)$ are $1$, meaning that the $\text{Tr}\, R \, USp(4)_g^2$ anomaly is equal to $-1$, which is the same as the contribution of six free fundamental chiral fields of $USp(4)_g$. In particular adding the fields above the one loop beta function will vanish. The global symmetry of the theory contains the $USp(4)\times U(1)_a\times U(1)_t$ symmetry coming from the $E_6$ SCFT. The $U(1)_t$ comes from the enlarged R-symmetry of the ${\cal N}=2$ superconformal algebra. Our assignment of charges is such that the moment map operators have $U(1)_t$ charge $+1$ while a dimension $d$ Coulomb branch operator has $U(1)_t$ charge $-d$.
  For the $U(1)_t$ not to be anomalous we assign charges $+\frac12$ to $q_L$ and $-1$ to $\phi_A$. We also have $SU(3)\times SU(6)\times U(1)_b$ coming from the extra fields we add. Under $U(1)_b$, the fields $q_L$ have charge $+1$ and $\phi_A$ charge $-1$.
 
\begin{figure}[t]
	\centering
  	\includegraphics[scale=0.42]{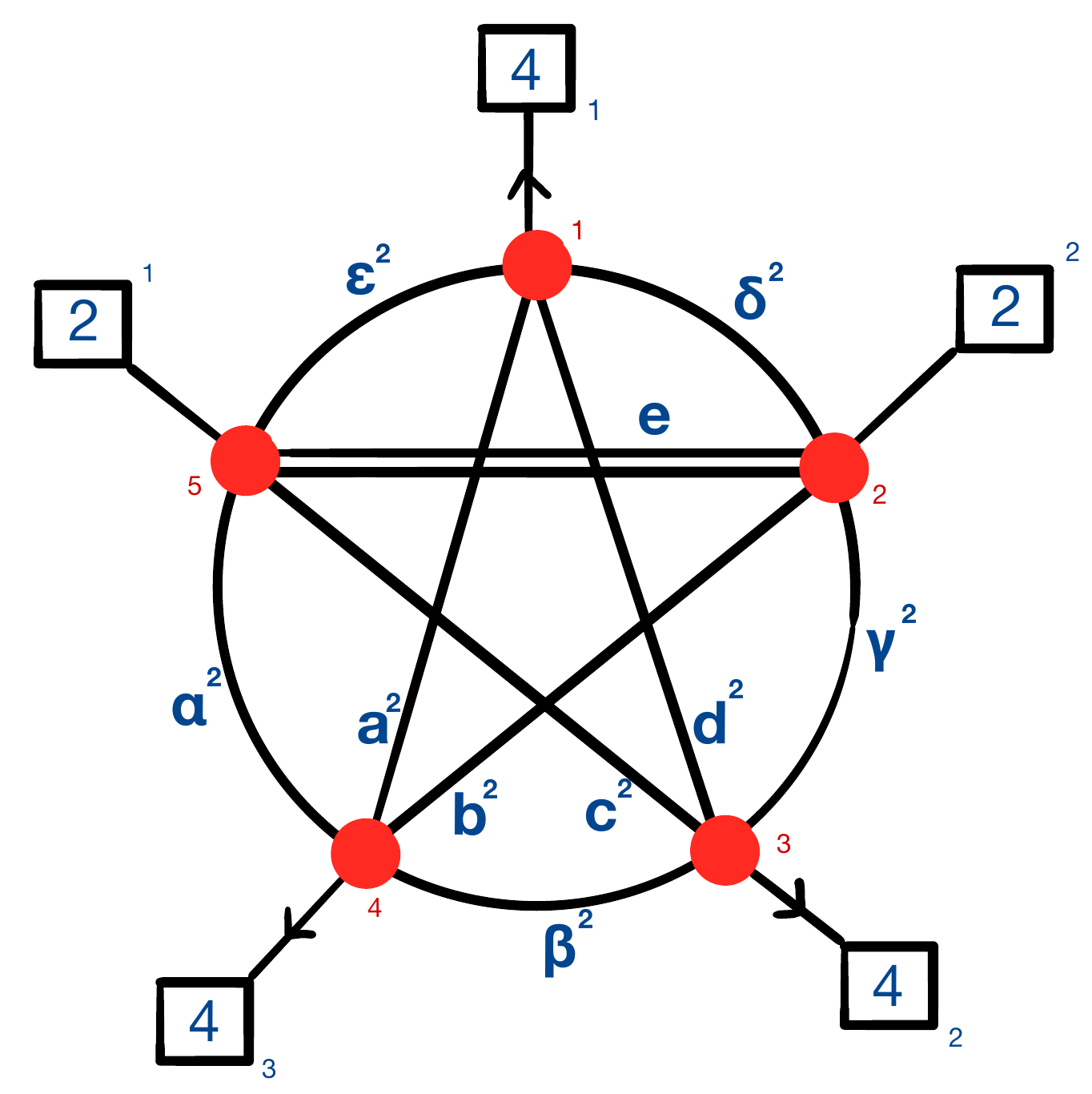}
    \caption{The quiver dual to $USp(4)_g$ gauging (with matter) of the $E_6$ MN SCFT. The red dots are $SU(2)$ gauge groups. The various letters denote fugacities for the ten abelian symmetries. The missing letters should be filled in by requiring the gauge symmetry to be non-anomalous. The theory has at the free point three $SU(4)$ symmetries and three $SU(2)$ symmetries. One of the three $SU(2)$ symmetries rotates the two bi-fundamental fields between gauge nodes $2$ and $5$. One needs to turn on the most general cubic gauge invariant superpotential. The theory is conformal as each $SU(2)$ gauge group has six flavors.}
    \label{F:satanicE6dual}
\end{figure}

 The $E_6$ SCFT has conformal anomalies, 
 \be
 a=\frac{41}{24}\,,\qquad c=\frac{13}6\,.
 \ee These are the anomalies which can be obtained from $5$  free vectors and $37$ free chiral fields using,
 \be
a=\frac3{16}\text{dim}\,{\frak G}+\frac1{48}\text{dim}\,{\frak R}\,,\qquad\qquad c=\frac1{8}\text{dim}\,{\frak G}+\frac1{24}\text{dim}\,{\frak R}\,,
\ee where $\text{dim}\,{\frak G}$ is the number of free vectors (dimension of the gauge group) and $\text{dim}\,{\frak R}$ is the number of free chiral superfields (dimension of the representation of the matter fields). We add to the model $10$ gauge fields of $USp(4)_g$ and additional six fundamentals and three two index traceless antisymmetric fields number of which is $39$.  The conformal anomalies of the theory are thus, $a=\frac{211}{48}$ and $c=\frac{121}{24}$. If we are after a conformal dual of this model it has to have,
\be
\text{dim}\,{\frak G}=5+10=15\,,\qquad \qquad \text{dim}\,{\frak R}=37+39=76\,.
\ee Having $\text{dim}\,{\frak G}=15$ and assuming the dual is a conformal Lagrangian theory, we have only two candidate gauge groups, $SU(4)$ and $SU(2)^5$. In fact we find a dual with the latter option. 
 We suggest that the theory has a dual description in terms of an $SU(2)^5$ conformal quiver gauge theory depicted in Figure \ref{F:satanicE6dual}. This model has $11$ bi-fundamental fields between various $SU(2)$ gauge groups and $16$ fundamentals of a single gauge group. This matter content amounts to $16\times 2+11\times 4=76$ free chiral fields, guaranteeing that the conformal anomalies match. Each $SU(2)$ gauge group has six flavors ensuring the one loop gauge beta functions vanish, and we soon verify that indeed both models have non-trivial conformal manifold. We will match the indices of the theories in expansions of fugacities. In particular, it will imply equality of the number of relevant and marginal operators.
 
 \

 \noindent {\bf The conformal manifold}
 
  The $E_6$ Minahan-Nemeschansky SCFT has moment map operators in the adjoint of $E_6$ which decompose into $USp(4)_g\times USp(4)\times U(1)_a$ as,
 
 \be
 {\bf 78} \to ({\bf 1},{\bf 1})_0\oplus ({\bf 4},{\bf 4})_{+3}\oplus ({\bf 4},{\bf 4})_{-3}\oplus ({\bf 5},{\bf 5})_0\oplus ({\bf 10},{\bf 1})_0\oplus ({\bf 1},{\bf 10})_0\,.
 \ee There are many marginal operators one can build and on a generic point of the conformal manifold all the symmetry is broken.  Let us denote the operators in $ ({\bf 4},{\bf 4})_{\pm3}$ as $M^{\pm}_{ij}$ and operators in $({\bf 5},{\bf 5})_0$ as $M_{ab}$. Then the marginal operators are,
 
 \be\label{margsE6}
 &&M^{\pm}_{ij}q^i_L\, (\, ({\bf 4},{\bf 6},{\bf 1})_{{\pm3},\frac32,1}\,)\,,\;\;\qquad  M_{ab}\phi^a_A\, (\, ({\bf 5},{\bf 1},{\bf 3})_{0,0,-1}\,)\,,\;\;\\
 && q^{(i}_{(L} q^{j)_a}_{M)} \phi_{aA}\, (\, ({\bf 1},{\bf 15},{\bf 3})_{0,0,1}\,)\,, \;\;\qquad \Phi_3 \, (\, ({\bf 1},{\bf 1},{\bf 1})_{0,-3,0}\,)\,.\nonumber
 \ee The operator $\Phi_3$ is the dimension three Coulomb branch operator of the $E_6$ SCFT and $({\bf X},{\bf Y},{\bf Z})_{q_a,q_t,q_b}$ denote representations under $(USp(4),SU(6),SU(3))_{U(1)_a,U(1)_t,U(1)_b}$. To compute the dimension of the conformal manifold we need to analyze the K\"ahler quotient $\{\lambda_I\}/G_{\C}$ \cite{Green:2010da} (see also \cite{Leigh:1995ep,Kol:2002zt}), where $\lambda_I$ are the marginal couplings and $G_\C$ is the complexified global symmetry group. In our case the couplings $\lambda_I$ are the ones for the operators in \eqref{margsE6} and $G_\C$ is $USp(4)\times SU(6)\times SU(3)\times U(1)_a\times U(1)_t\times U(1)_b$.
  The K\"ahler quotient is not empty. For example $M\phi$ times $q^2 \phi$ is not charged under any $U(1)$s and contains a component  in $({\bf 5},{\bf 15},{\bf 6})$. Taking it to symmetric sixth power we get singlet of all the symmetries.  This deformation breaks the $U(1)_b$ symmetry. Also the $M \phi$ coupling breaks the $USp(4)\sim SO(5)$ symmetry\footnote{We will not be careful with the global structure of the groups in this note.}  to its $SO(2)\times SO(3)$ subgroup, the $SU(3)$ to its $SO(3)$, and furthermore locks the two $SO(3)$ groups to the diagonal.  The $SU(6)$ is broken by the operators $q^2\phi$ as follows $SU(6)\rightarrow SU(2)\times SU(3)\rightarrow U(1)\times SU(3)$, where the first arrow uses the embedding of the symmetry such that ${\bf 6}_{SU(6)}\rightarrow {\bf 2}_{SU(2)} {\bf 3}_{SU(3)}$ and in the second the $SU(2)$ is broken to its Cartan. This $SU(3)$ and the one acting on the antisymmetric are then locked to the diagonal. The combined effect of both of them is to break $USp(4)\times SU(6)\times SU(3)\times U(1)_b$ to $SO(3) \times U(1)^2$. There is a 1d subspace that preserves the $SO(3) \times U(1)^2 \times U(1)_t\times U(1)_a$ symmetry though a generic choice of these operators also breaks the $SO(3)$, spanning an 8d subspace preserving only $U(1)^2 \times U(1)_t\times U(1)_a$. Finally, we can turn on the rest of the marginal operators, $\Phi_3$ and $M^{\pm}q$, which can be used to break all $U(1)$ symmetries as well. This gives a $53$ dimensional conformal manifold on a generic point of which no symmetry is preserved. 
 
\begin{figure}[t]
	\centering
  	\includegraphics[scale=0.42]{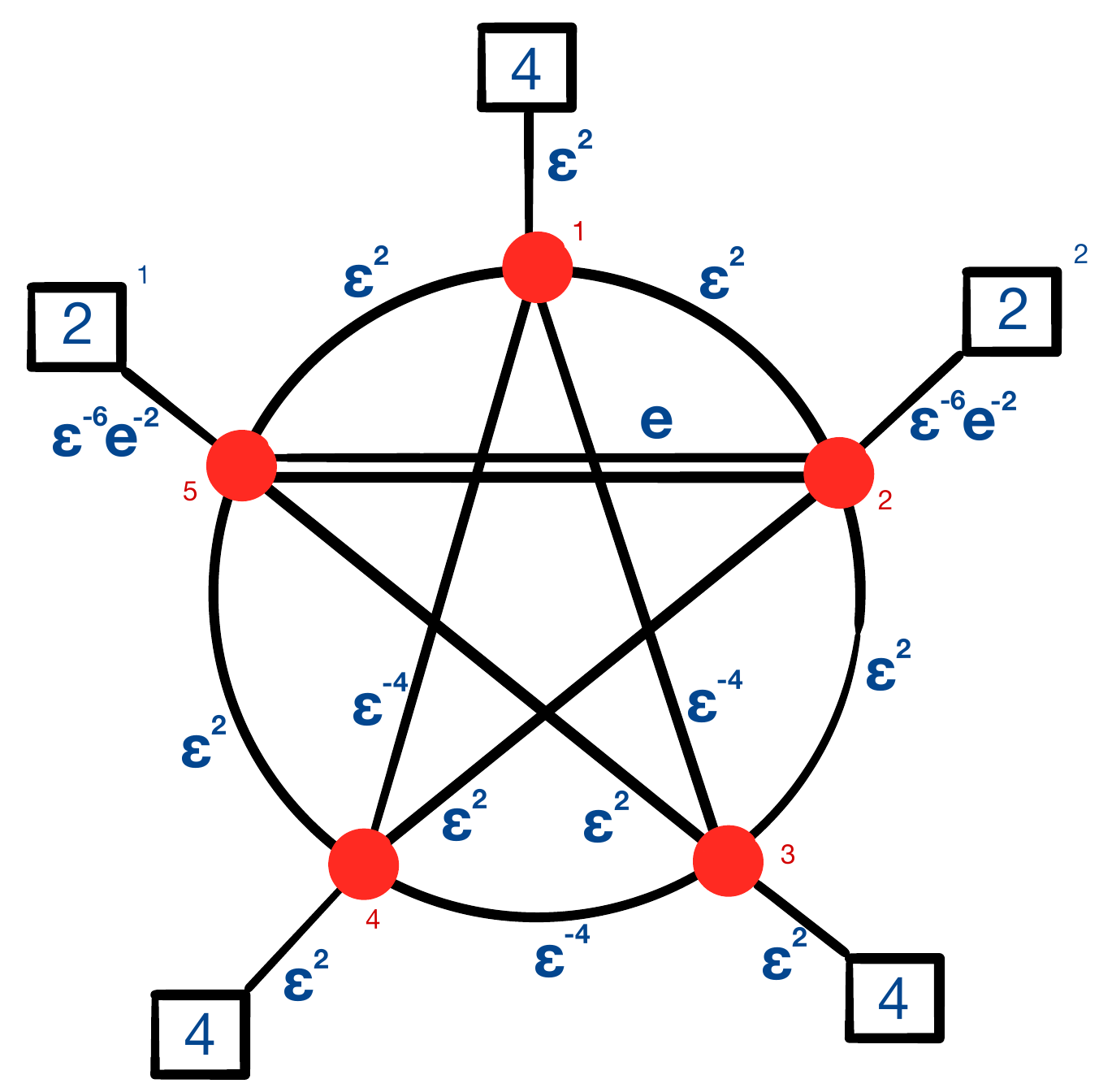}
    \caption{Going on the conformal manifold we necessarily break some of the symmetry. A  sublocus of the conformal manifold which is easy to identify is the one on which the ten abelian symmetries are broken to $U(1)_\epsilon\times U(1)_e$ denoted on the quiver. The three $SU(4)$ symmetries are broken to diagonal $USp(4)$. The three $SU(2)$ symmetries are not broken.  One turns on superpotentials consistent with the charges in the figure.}
    \label{F:satanicdefE6dual}
\end{figure}

 Let us analyze the conformal manifold on the quiver side. We have ten anomaly free abelian symmetries, which we denote as $U(1)_{a,b,c,d,e}$ and $U(1)_{\alpha,\beta,\gamma,\delta,\e}$ (see Figure \ref{F:satanicE6dual}), and non-abelian symmetry $SU(4)^3\times SU(2)^3$ at the free point. We have many marginal deformations and let us first list the operators which do not transform under $SU(2)^3$ by detailing their charges,
 
 \be\label{quivoperators}
&& A_{13}:\; {\bf 4}_1\otimes {\bf 4}_3 \times \frac1{bd\alpha\beta\e\delta} \,,\qquad
  A_{12}:\; {\bf 4}_1\otimes {\bf 4}_2 \times \frac1{ac\beta\gamma\e\delta} \,,\qquad
   A_{23}:\; {\bf 4}_2\otimes {\bf 4}_3 \times \frac1{abcd\alpha\gamma} \,,\qquad\\
&& M_\e:\; c^2 d^2\e^2\,,\qquad    M_\delta:\; a^2 b^2\delta^2\,,\qquad    
M_\beta:\; a^2 d^2\beta^2\,,\qquad  M_b:\; \gamma^2 \beta^2 b^2\,,\qquad \nonumber\\
&& M_c:\; \alpha^2 \beta^2 c^2\,,\qquad           M_a:\; \alpha^2 \epsilon^2 a^2\,,\qquad      M_d:\; \gamma^2 \delta^2 d^2\,.\nonumber
 \ee The   $A_{ij}$ are cubic operators winding between the $i$th and $j$th $SU(4)$ group, while the $M_\#$ operators are cubic operators corresponding to triangles in the quiver. For the latter case, when $\#$ is a Greek letter then these are triangles containing one bi-fundamental running along the circle (denoted by the Greek letter $\#$) and two internal ones, while when $\#$ is a Latin letter, then these are triangles containing one internal bi-fundamental (denoted by the Latin letter $\#$) and two circle ones. We immediately note that, \be\label{simpledef}(A_{13}A_{12}A_{23})^4 (M_\e M_\delta M_a M_b M_c M_d)^2\,,\ee is not charged under any abelian symmetries, does not transform under $SU(2)^3$, and  also contains an invariant of the three $SU(4)$ symmetries if we contract the $SU(4)$ indices with the epsilon symbols. Thus the conformal manifold is not empty. The effect of these operators is to break all abelian symmetries, save for $U(1)_e$, down to a single one which we denote, by abuse of notation,  as $U(1)_{\epsilon}$ (see Figure \ref{F:satanicdefE6dual} for their charges in terms of $U(1)_\epsilon$). Furthermore, the $SU(4)$ groups are all locked together and further broken. The minimal possible breaking of the $SU(4)$ groups is either to $USp(4)$ or $SO(4)$, both happen along a $1d$ subspace. A generic combination also break these symmetries to the Cartan. This gives a $3d$ subspace along which a $U(1)^2 \times U(1)_{\epsilon} \times U(1)_e \times SU(2)^3$ global symmetry is preserved.
 
 Let us continue to study the conformal manifold by going along the $1d$ subspace preserving the $USp(4)$, turning the marginal operators charged under the $SU(2)$ symmetries and only considering their charges under the symmetries preserved on this submanifold. We have the triplet of operators, which we denote as $M_j^{i=1,2,3}$, and carry the charges: ${\bf 2}_j\otimes {\bf 4}\times \frac1{e^2\e^2}$. These are the operators running between the $i$th $SU(2)$ group and $j$th $SU(4)$ group. We have three operators $M^{i=1,2,3}$ charged ${\bf 2}\times \e^4 e$ corresponding to three triangles including bi-fundamentals transforming under one the $SU(2)$ groups.  Finally we have an operator $M_0$ charged ${\bf 2}\otimes {\bf 2}_1\otimes {\bf 2}_2\times \frac1{e^3\e^{12}}$ which corresponds to an operator winding between the two $SU(2)_i$ groups. Note that it is easy to build invariants here. For example,  $(M_0)^2$ is a singlet of $SU(2)_1\times SU(2)_2$ (when we contract  the indices with $\epsilon$ symbol) and is in the adjoint of $SU(2)$ and has charge $\frac1{e^6\e^{24}}$, while say $(M^1 M^2)^3$ contains an adjoint of $SU(2)$ and has charge $e^6\e^{24}$. Thus, contracting the two combinations we get a singlet. This deformation breaks $SU(2)_1\times SU(2)_2$, at least to the diagonal combination, breaks $SU(2)$ completely and identifies  $e=\e^{-4}$. In particular $M^j$ and $M_0$, under the preserved symmetry, are in the $6\times {\bf 1}\oplus 2\times ({\bf 3}\oplus {\bf 1})$, while the broken currents are in $4\times {\bf 1}\oplus 2\times {\bf 3}$ meaning that we get a five dimensional submanifold preserving $USp(4)\times SU(2)_{diag}\times U(1)_\e$. We also have an operator in the adjoint of $SU(2)_{diag}$ which breaks it to the Cartan if turned on.

We can continue turning on marginal operators and breaking the symmetry further. The operator $M^i_j$ now are charged ${\bf 2}\otimes {\bf 4} \times \e^6$ while $M_\beta$ is charged $\e^{-12}$. In particular say taking $M_1^1 M_1^2 M_\beta$ is a singlet of all the remaining symmetries. These operators break the $U(1)_\epsilon$ but preserve an $SU(2)_{diag}\times SU(2)'$. Here we decompose $USp(4)$ to $SU(2)_{diag}\times SU(2)'$ such that ${\bf 4} \to {\bf 2}_{diag}+{\bf 2}'$. Some components of the operators $M^i_j$ will recombine with the conserved currents, some will contribute exactly marginal operators in the singlet of $SU(2)_{diag}\times SU(2)'$, and we will also get several operators in ${\bf 2}_{diag}\otimes {\bf 2}'$. Turning on these we can break $SU(2)_{diag}\times SU(2)'$ to a diagonal $SU(2)$ and get several marginal operators in adjoint of it.
 Turning on one of the adjoints we can break the symmetry to the Cartan while turning on the rest we completely break the symmetry.
  All in all we break the symmetry completely on the conformal manifold. Thus the dimension of the manifold is the number of marginal operators minus the currents which gives us  $53$ dimensional conformal manifold.
  
  \

\noindent {\bf The supersymmetric index}

\noindent The index in both duality frames is given by (for definitions of the index see Appendix \ref{app:schur}),
 
 \be
&& 1+32 (qp)^{\frac23}+53 qp+31(qp)^{\frac23} \left(q+p\right)+586 (qp)^{\frac43}+48qp \left(q+p\right)+\\
&&1463(qp)^{\frac53}+ 31(qp)^{\frac23}\left(q^2+p^2\right)+1058(qp)^{\frac43} \left(q+p\right)+\cdots\,.\nonumber
 \ee  On the $E_6$ side of the duality the index can be computed using either the construction of \cite{Gadde:2010te,Gadde:2015xta} or the Lagrangian of \cite{Zafrir:2019hps}. Moreover, as on the quiver side we have a rank five gauge theory making the evaluation of the index computationally intense, one can take the Schur limit of the index, even though the theory is only ${\cal N}=1$, to simplify computations. The limit is $p^2=q$ \cite{Razamat:2019vfd}\footnote{We thank C. Beem and C. Meneghelli for pointing out to us this relation to  the Schur index.} and then one can use the expressions for the index of $E_6$ SCFT using Schur polynomials \cite{Gadde:2011ik,Gadde:2011uv},
 
\be
I_{E_6}({\bf z}_1,{\bf z}_2,{\bf z}_3) =\frac{1}{(1-q)^2(1-q^2)(q;q)^4\prod_{i\neq j}\prod_{l=1}^3 (qz^{(l)}_i/z^{(l)}_j;q) }\sum_{\lambda_1=0}^\infty \sum_{\lambda_2=0}^{\lambda_1}
\frac{\prod_{l=1}^3\chi_{\lambda_1,\lambda_2}({\bf z}^{(l)})}{\chi_{\lambda_1,\lambda_2}(q,1,q^{-1})}\,.\nonumber\\
\ee Here ${\bf z}_i$ are the fugacities for the  $SU(3)^3$ maximal subgroup of $E_6$, $\lambda_1$ and $\lambda_2$ are the lengths of the Young tableaux defining representations of $SU(3)$, and $\chi_{\lambda_1,\lambda_2}$ are the corresponding Schur polynomials.
 Then we define the single letter partition function of the extra fields on the $E_6$ side of the duality to be,

\be
M_A(z_1,z_2;q)=\frac{q^{\frac12}}{1-q}\left(6\chi_{\bf 4}(z_1,z_2)+3\chi_{\bf 5}(z_1,z_2)\right)-\left(\frac{q}{1-q}+\frac{q^{\frac12}}{1-q^{\frac12}}\right)\chi_{\bf 10}(z_1,z_2)\,,
\ee giving the index,

\be
&&I_A = \oint\frac{dx_1}{2\pi i x_1} \oint\frac{dx_2}{2\pi i x_2} \Delta_{USp(4)}(x_1,x_2)\times\,\\
&&\;\;\;\;\; I_{E_6}\left(\frac{x_1^{\frac23}}{x_2^{\frac13}},\frac{x_2^{\frac23}}{x_1^{\frac13}},\frac1{x_1^{\frac13}x_2^{\frac13}};\frac{x_1^{\frac23}}{x_2^{\frac13}},\frac{x_2^{\frac23}}{x_1^{\frac13}},\frac1{x_1^{\frac13}x_2^{\frac13}};\frac1{x_1^{\frac13}x_2^{\frac13}},\frac1{x_1^{\frac13}x_2^{\frac13}},x_1^{\frac23}x_2^{\frac23}\right) PE\left[M_A(x_1,x_2;q)\right]\,.\nonumber
\ee  Here by $\Delta_G({\bf z})$ we denote the $G$ invariant, Haar, measure.
On the quiver side of the duality the contribution of the matter is,

\be
&&M_B(z_1,z_2,z_3,z_4,z_5;q)=-\left(\frac{q}{1-q}+\frac{q^{\frac12}}{1-q^{\frac12}}\right)\sum_{i=1}^5\chi_{\bf 3}(z_i)+\\
&&\;\;\; \frac{q^{\frac12}}{1-q}\left(\sum_{i=1}^5\chi_{\bf 2}(z_i)(\chi_{\bf 2}(z_{i+1})+\chi_{\bf 2}(z_{i+2})+4)+(\chi_{\bf 2}(z_2)-2)(\chi_{\bf 2}(z_5)-2)-4\right)\,,\nonumber
\ee with the index given by,

\be
I_B =\prod_{i=1}^5 \left[\oint\frac{dz_i}{2\pi i z_i} \Delta_{SU(2)}(z_i)\right] \, PE\left[M_B(z_1,z_2,z_3,z_4,z_5;q)\right]\,.
\ee Both indices can be evaluated to rather high order to give $I_A=I_B$, and explicitly,

\be
&&1+32 q+84 q^{\frac32}+696 q^2+2648 q^{\frac52}+13267 q^3+51379 q^{\frac72}+ 209576 q^4+765123 q^{\frac92}+\;\;\;\\
&&\;\;\;\;\; 2769413 q^5+9428456 q^{\frac{11}2}+31348364 q^{6}+\cdots\,.\nonumber
\ee

We thus have compelling evidence that in fact the $USp(4)_g$ gauging of the $E_6$ MN theory is conformally dual to the ${\cal N}=1$ quiver theory.

\

\section{Dual of $E_7$ MN theory with $SU(2)$ subgroup gauged}

Our second example of a duality has on one side  an $\mathcal{N}=1$ conformal gauge theory with a weak coupling limit, while the other contains an intrinsically strongly interacting part, which here is the rank $1$ $E_7$ MN theory. The gauge theory side has gauge group $USp(4)$, three chiral fields in the traceless second rank antisymmetric representation and twelve chiral fields in the fundamental representation. With this matter content the one loop beta function vanishes. The theory has a non-anomalous global symmetry of $U(1)_t\times SU(3)\times SU(12)$. Under the $U(1)_t$ symmetry the antisymmetric fields have charge $-1$ and the fundamental fields have charge $+\frac12$. The model has classically marginal operators made from a contraction of the antisymmetric and two fundamental chirals. This marginal operator is in the $(\bold{3}, \bold{66})$ of $SU(3)\times SU(12)$ and is uncharged under $U(1)_t$. As we shall show there is a non-trivial K\"ahler quotient, and so by the arguments of \cite{Leigh:1995ep,Green:2010da}, it exists as an SCFT with a conformal manifold containing the weak coupling point. It is possible to show that the $SU(3)\times SU(12)$ can be completely broken on the conformal manifold leading to a $3 \times 66 - 143 - 8 = 47$ dimensional conformal manifold, on a generic point of which only the $U(1)_t$ is preserved. The theory has $72$ chiral operators of dimension $2$ given by the symmetric invariant of the antisymmetric chiral fields, transforming in the $\bold{6}$ of $SU(3)$ and with charge $-2$ under $U(1)_t$, and the antisymmetric invariant of the fundamental chiral fields, transforming in the $\bold{66}$ of $SU(12)$ and with charge $+1$ under $U(1)_t$. 

The dual side is an $\mathcal{N}=1$ $SU(2)$ gauging of the $\mathcal{N}=2$ rank one SCFT with $E_7$ global symmetry with four chiral fields in the doublet representation for the $SU(2)$. As the $E_7$ SCFT provides an effective number of eight chiral doublets for the $SU(2)$ beta function \cite{Argyres:2007ws}, the latter vanishes. The theory has a $U(1)_t \times SU(4)\times SO(12)$ global symmetry. Here the $SU(4)$ is the symmetry rotating the $SU(2)$ doublets and $SO(12)$ is the commutant of $SU(2)$ inside $E_7$. The abelian symmetry is the anomaly free combination of the $U(1)$ acting on the four $SU(2)$ doublets and $U(1)_t$, which is the commutant of the $\mathcal{N}=1$ $U(1)_R$ in the $\mathcal{N}=2$ $U(1)_R \times SU(2)_R$. Using the duality in \cite{Argyres:2007ws}, it is straightforward to show that under this symmetry, which by abuse of notation we will denote $U(1)_t$, the $SU(2)$ doublets have charge $-1$ where we have normalized $U(1)_t$ as before,  such that the moment map operators of the $E_7$ SCFT have charge $+1$.

We have relevant operators of dimension two given by the moment maps of the $E_7$ SCFT which transform in the $\bold{133}_{E_7}$. After the gauging these decompose to $SU(2) \times SO(12)$ according to $\bold{133}_{E_7} \rightarrow \bold{3}_{SU(2)} + \bold{2}_{SU(2)}\bold{32}_{SO(12)} + \bold{66}_{SO(12)}$. In particular, we have the gauge variant $\bold{2}_{SU(2)}\bold{32}_{SO(12)}$ operators, which can be made into a dimension three gauge invariant operators via a contraction with the $SU(2)$ chiral doublets. This gives a classically marginal operator in the $\bold{4}_{SU(4)}\bold{32}_{SO(12)}$. Additionally, as the moment map operators carry charge $+1$ under the non-anomalous $U(1)_t$ and the chiral $SU(2)$ doublets carry charge $-1$, it is uncharged under $U(1)_t$. As we shall show there is a non-trivial K\"ahler quotient, and so again it follows that this theory exists as an SCFT with a conformal manifold containing the weak coupling point of the $SU(2)$. It is possible to show that the $SU(4)\times SO(12)$ global symmetry can be completely broken on the conformal manifold leading to a $4 \times 32 - 66 - 15 = 47$ dimensional conformal manifold, on a generic point of which only $U(1)_t$ is preserved. The theory has $72$ dimension two operators, $66$ of which are given by the moment map operators associated with the $SO(12)$ and are in the $\bold{66}$ of $SO(12)$ and have $U(1)_t$ charge $+1$. The remaining $6$ operators come from the antisymmetric invariant of the $SU(2)$ doublets, transform in the $\bold{6}$ of $SU(4)$ and carry charge $-2$ under $U(1)_t$. Note that as the $SU(4)$ did not originate from an $\mathcal{N}=2$ theory, it does not have moment map operators.

The $E_7$ SCFT has conformal anomalies, 
 \be
 a=\frac{59}{24}\,,\qquad c=\frac{19}6\,.
 \ee These are the anomalies which can be obtained from $7$  free vectors and $55$ free chiral fields. We add to the model the $3$ gauge fields associated with the $SU(2)$ gauge group and four chiral fields in the doublet representation of $SU(2)$, giving $8$ extra chiral fields. The conformal anomalies of the theory are thus, $a=\frac{51}{16}$ and $c=\frac{31}{8}$. If we are after a conformal dual of this model it has to have,
\be
\text{dim}\,{\frak G}=3+7=10\,,\qquad \qquad \text{dim}\,{\frak R}=55+8=63\,.
\ee Having $\text{dim}\,{\frak G}=10$ and assuming the dual is a conformal Lagrangian theory we have only one candidate gauge group, $USp(4)$, and we find such a dual mentioned above.
 This model has $12$ fundamental fields and three tracelss two index antisymmetric fields. This matter content amounts to $12\times 4+5\times 3=63$ free chiral fields, guaranteeing that the conformal anomalies match.

So far we have seen that both theories exist as interacting SCFTs with a conformal manifold and have the same conformal anomalies. We have also seen that the dimension of the conformal manifold, generically preserved global symmetry and relevant operators all match between the two theories. This prompts us to propose that these two theories are in fact dual and share the same conformal manifold. The global symmetry at the weak coupling point differs, but this can easily be accounted for as most of the global symmetry is broken when moving on the conformal manifold. The $U(1)_t$ symmetry is the only part that is never broken and so must match between the two theories. We next present evidence for our claim.

\subsection*{Anomalies}

We begin by comparing the 't Hooft anomalies of the two theories. Only anomalies for symmetries that are preserved along a path on a conformal manifold connecting the two theories must match. Furthermore, only the flavor $U(1)_t$ and $U(1)_R$ are preserved on generic points and so these must match and we shall compare only these for now. The $\mathcal{N}=1$ $USp(4)$ gauge theory contains $10$ vector multiplets, $48$ chiral fields with $U(1)_t$ charge $\frac12$ and free R-charge, and $15$ chiral fields with $U(1)_t$ charge $-1$ and free R-charge. From this data, all anomalies involving the flavor $U(1)_t$ and $U(1)_R$ can be calculated.

For the dual side, it is convenient to use the duality of \cite{Argyres:2007ws}. It implies that the $\mathcal{N}=2$ rank one $E_7$ SCFT has the same anomalies as $7$ vector multiplets, $48$ chiral fields with $U(1)_t$ charge $\frac12$ and free R-charge, and $7$ chiral fields with $U(1)_t$ charge $-1$ and free R-charge. Additionally, we have the $SU(2)$ with the four chiral doublets which contributes $3$ vector multiplets and $8$ chiral fields with $U(1)_t$ charge $-1$ and free R-charge. Overall we find the same effective matter content as the $\mathcal{N}=1$ $USp(4)$ gauge theory and so all anomalies involving $U(1)_t$ and $U(1)_R$ will match.

\subsection*{Superconformal index}

\noindent We can next match the superconformal index. It is not hard to compute it for the $\mathcal{N}=1$ $USp(4)$ gauge theory finding,
\bea
I & = & 1 + (p q)^{\frac{2}{3}}(6 t^{-2} + 66 t) + 46 p q + (p q)^{\frac{2}{3}}(p+q)(6 t^{-2} + 66 t ) + (p q)^{\frac{4}{3}}(21 t^{-4} + 279 t^{-1} + 2016 t^2) \nonumber \\ & + & p q (p+q) (45-t^{-3}) + (p q)^{\frac{2}{3}}(p^2+pq+q^2)(6 t^{-2} + 66 t ) + (p q)^{\frac{5}{3}}(159 t^{-2} + 1356 t) + ...  \label{Ind}
\eea
Here we use $t$ for the $U(1)_t$ fugacity and we have only refined with respect to symmetries that are preserved generically on the conformal manifold. For the dual side, we utilize the index of the $E_7$ SCFT computed in \cite{Agarwal:2018ejn}. Using it we find result exactly matching with (\ref{Ind}). Here also we can compute the superconfomal index in the Schur limit on both sides to high order in an expansion in fugacities. To compute the Schur index we set $t=1$ and $q=p^2$ as before\footnote{Note that since $U(1)_t$ is preserved on the conformal manifold one can utilize various  limits of the index discussed in \cite{Gadde:2011uv}. Let us here comment on the Coulomb limit. It is convenient to assign R charge $0$ to chiral fields with $U(1)_t$ charge $\frac12$ and R-charge $2$ to fields with $U(1)_t$ charge $-1$. This assignment is non anomalous. The Coulomb limit is $pq/t\to x$, while $p,q,t\to 0$ in the notations of this footnote, and it is easy to compute. On the $E_7$ side the $E_7$ SCFT  contributes a factor of $1/(1-x^4)$ coming from the dimension four Coulomb branch operator, while the $SU(2)$ gauging contributes $\oint \frac{dz}{2\pi ii z} \Delta_{SU(2)}(z)\frac1{(1-x z^{\pm1})^4}=\frac{1-x^4}{(1-x^2)^6}$. On the gauge theory side the only contributions come from fields in the ${\bf 5}$ and we have $\oint\frac{dz_1}{2\pi i z_1} \oint\frac{dz_2}{2\pi i z_2} \Delta_{USp(4)}(z_1,z_2)\frac1{((1-x)(1-x z_1^{\pm1}z_2^{\pm1}))^3}=\frac1{(1-x^2)^6}$. The two dual indices manifestly and non-trivially match.}.
  To compute this result we first use \cite{Gadde:2011ik,Gadde:2011uv} to write the Schur index of the $E_7$ SCFT as,
\be\label{IndE7}
I_{E_7}({\bf z}^{(1)},{\bf z}^{(2)},a) &=&\frac{(qa^{\pm2};q)^{-1}(q^2a^{\pm2};q)^{-1}}{(1-q)(1-q^2)^2(1-q^3)(q;q)^6\prod_{i\neq j}\prod_{l=1}^2 (qz^{(l)}_i/z^{(l)}_j;q) }\\
&&\;\;\; \sum_{\lambda_1=0}^\infty \sum_{\lambda_2=0}^{\lambda_1}\sum_{\lambda_3=0}^{\lambda_2}
\frac{\chi_{\lambda_1,\lambda_2,\lambda_3}(q^{\frac12}a,q^{\frac12}a^{-1},q^{-\frac12}a,q^{-\frac12}a^{-1})\prod_{l=1}^2\chi_{\lambda_1,\lambda_2,\lambda_3}({\bf z}^{(l)})}{\chi_{\lambda_1,\lambda_2,\lambda_3}(q^{\frac32},q^{\frac12},q^{-\frac12},q^{-\frac32})}\,.\nonumber
\ee Here ${\bf z}^{(i)}$ are fugacities for two $SU(4)$ symmetries and $a$ is a fugacity for an $SU(2)$. The $SU(2)$ appears in the decomposition of $E_7$ to $SO(12)\times SU(2)$, while  the two $SU(4)\sim SO(6)$ appear in the decomposition of $SO(12)\to SO(6)\times SO(6)$. The integers $\lambda_i$ label the Young tableaux associated to representations of $SU(4)$, and $\chi_{\lambda_1,\lambda_2,\lambda_3}$ are the Schur polynomials for $SU(4)$.
Then we define the single letter partition function of the extra fields on the $E_7$ side of the duality to be,
\be\label{MAE7}
M_A(a;q)=\frac{q^{\frac12}}{1-q}\left(4\chi_{\bf 2}(a)\right)-\left(\frac{q}{1-q}+\frac{q^{\frac12}}{1-q^{\frac12}}\right)\chi_{\bf 3}(a)\,,
\ee giving the index,
\be
&&I_A = \oint\frac{dz}{2\pi i z}  \Delta_{SU(2)}(z)I_{E_7}\left({\bf 1},{\bf 1},z\right) PE\left[M_A(z;q)\right]\,.\nonumber
\ee On the quiver side of the duality the contribution of the matter is,
\be\label{MBE7}
&&M_B(z_1,z_2;q)=\frac{q^{\frac12}}{1-q}\left(12\chi_{\bf 4}(z_1,z_2)+3\chi_{\bf 5}(z_1,z_2)\right)-\left(\frac{q}{1-q}+\frac{q^{\frac12}}{1-q^{\frac12}}\right)\chi_{\bf 10}(z_1,z_2)\,,\nonumber
\ee with the index given by,
\be
I_B =\oint\frac{dz_1}{2\pi i z_1} \oint\frac{dz_2}{2\pi i z_2} \Delta_{USp(4)}(z_1,z_2)\, PE\left[M_B(z_1,z_2;q)\right]\,.
\ee In both duality frames we obtain that it is equal to,
\be
&&I_A=I_B=1+72 q+118 q^{\frac32}+2504 q^2+6625 q^{\frac52}+60894 q^{3}+188762 q^{\frac72}+1157937 q^{4}+\nonumber\\
&&3722096 q^{\frac94}+18018345 q^{5}+57271940 q^{\frac{11}2}+236762366 q^{6}+731094087 q^{\frac{13}2}+\\
&&\,\,\,2694503918 q^{7}+8036370246 q^{\frac{15}2}+27107273596 q^{8}+\cdots\,.\nonumber
\ee

\

\subsection*{Structure of the conformal manifold} 

Finally, we can study the structure of the conformal manifold in more detail. Specifically, we consider whether it may be possible to connect the two theories through a path in the conformal manifold preserving more than the $U(1)$ flavor symmetry. For this we need to better examine the conformal manifold of the two theories. We shall start with the frame with the $E_7$ SCFT. Here the marginal operators are in the $\bold{4}_{SU(4)}\bold{32}_{SO(12)}$ of the $SU(4) \times SO(12)$ global symmetry. First, as the $\bold{32}_{SO(12)}$ has a non-trivial quartic fully antisymmeric invariant, there is at least one exactly marginal combination. Say we insert it into the superpotential, then the symmetry would be reduced to the subgroup keeping that element fixed, that is to a subgroup of $SU(4) \times SO(12)$ under which the $\bold{4}_{SU(4)}\bold{32}_{SO(12)}$ contains a singlet.

Going over the list of subgroups, we find the following solution. We break $SO(12)$ to its $SU(2)\times USp(6)$ subgroup such that $\bold{32}_{SO(12)} \rightarrow \bold{4}_{SU(2)} + \bold{2}_{SU(2)} \bold{14}_{USp(6)}$, $SU(4)$ to its $SU(2)$ subgroup such that $\bold{4}_{SU(4)}\rightarrow \bold{4}_{SU(2)}$ and we identify the two $SU(2)$ factors. Under this breaking we have that:

\bea
\bold{4}_{SU(4)}\bold{32}_{SO(12)}\rightarrow 1 + \bold{3}_{SU(2)} + \bold{5}_{SU(2)} + \bold{7}_{SU(2)} + (\bold{3}_{SU(2)} + \bold{5}_{SU(2)}) \bold{14}_{USp(6)}, \label{MSP}
\eea   
and there is indeed a singlet. Additionally the conserved currents of $SU(4) \times SO(12)$ decompose as:

\bea
\bold{15}_{SU(4)} &\rightarrow& \bold{3}_{SU(2)} + \bold{5}_{SU(2)} + \bold{7}_{SU(2)}, \\ \bold{66}_{SO(12)} &\rightarrow& \bold{3}_{SU(2)} + \bold{21}_{USp(6)} + \bold{3}_{SU(2)}\bold{14}_{USp(6)}. 
\eea

As $SU(4) \times SO(12)$ is broken to $SU(2)\times USp(6)$ by the deformation, the additional conserved currents must be eaten by marginal operators. Examining (\ref{MSP}), we see that we indeed have superpotential terms with the correct charges to merge with the conserved currents to form long multiplets. These superpotential terms then become marginally irrelevant and so we are left with $1+\bold{5}_{SU(2)} \bold{14}_{USp(6)}$ as the marginal operators. This suggests that there is a $1d$ subspace on the conformal manifold along which the preserved symmetry is $U(1)_t\times SU(2)\times USp(6)$. Along that subspace, we have $70$ additional marginal operators in the $\bold{5}_{SU(2)} \bold{14}_{USp(6)}$. The relevant dimension two operators carry charges of

\be
(1+\bold{5}_{SU(2)}) t^{-2} + (\bold{3}_{SU(2)} + \bold{21}_{USp(6)} + \bold{3}_{SU(2)}\bold{14}_{USp(6)})\,t,
\ee  
under the preserved $U(1)_t\times SU(2)\times USp(6)$ global symmetry.

Next we turn to the $\mathcal{N}=1$ $USp(4)$ gauge theory. Here the marginal operators are in the $(\bold{3}, \bold{66})$ of the $SU(3)\times SU(12)$ global symmetry. We can again show that there is $1d$ subspace along which the $SU(3)\times SU(12)$ global symmetry is broken to $SU(2)\times USp(6)$. For this we consider the embedding $SU(2)\times USp(6) \subset SO(12) \subset SU(12)$, and $SO(3)\subset SU(3)$ and take the diagonal $SU(2)$. Under this subgroup we have that:

\bea
\bold{3}_{SU(3)}\bold{66}_{SU(12)}\rightarrow 1 + \bold{3}_{SU(2)} + \bold{5}_{SU(2)} + \bold{3}_{SU(2)}\bold{21}_{USp(6)} + (1 + \bold{3}_{SU(2)} + \bold{5}_{SU(2)}) \bold{14}_{USp(6)}, \nonumber \\ \label{MSP1}
\eea
and there is indeed a singlet. We next need to consider the operators eaten by the broken currents, for which we need to consider the decomposition of the $SU(3)\times SU(12)$ conserved currents:

\bea
\bold{8}_{SU(3)} &\rightarrow& \bold{3}_{SU(2)} + \bold{5}_{SU(2)}, \\ \bold{143}_{SU(12)} &\rightarrow& \bold{3}_{SU(2)} + (1+\bold{3}_{SU(2)})(\bold{21}_{USp(6)}+\bold{14}_{USp(6)}). 
\eea

Again we find that we have sufficient superpotential terms to eat the broken currents, and we are left with: $1+\bold{5}_{SU(2)} \bold{14}_{USp(6)}$. Thus, we see that we indeed find a $1d$ subspace along which a $U(1)_t\times SU(2)\times USp(6)$ global symmetry is preserved\footnote{We can continue and break the symmetry completely by turning on the marginal operator in $\bold{5}_{SU(2)} \bold{14}_{USp(6)}$. In particular turning on this operator we can preserve along an additional $1d$ locus a diagonal combination of the $SU(2)$ and $SU(2)$ subgroup of $USp(6)$   such that ${\bf 6}_{USp(6)}\to {\bf 6}_{SU(2)}$. Doing so the remaining marginal operators are in the $1+{\bf 13}+2\times{\bf 9} +{\bf 7}+2\times {\bf 5} $ of the preserved $SU(2)$. Indeed we have a singlet and we can continue to break the $SU(2)$ further to the Cartan and then completely. All in all in the end we obtain $47$ dimensional conformal manifold.}.
 Furthermore, the remaining marginal operators match those found in the other frame. Finally we note that the relevant dimension two operators carry charges of

\be
(1+\bold{5}_{SU(2)}) t^{-2}+(\bold{3}_{SU(2)} + \bold{21}_{USp(6)} + \bold{3}_{SU(2)}\bold{14}_{USp(6)})\,t\, ,
\ee  
under the preserved $U(1)_t\times SU(2)\times USp(6)$ global symmetry. These indeed match those found in the other frame. Moreover it is also possible to check that the Schur indices refined with the $SU(2)\times USp(6)$ fugacities agree in an expansion in $q$. To perform this computation one should change the $4$ in \eqref{MAE7} to $\chi_{\bf 4}(u)$ of $SU(2)_u$, the $12$ and the $3$ in \eqref{MBE7} to $\chi_{\bf 2}(u)\times \chi_{\bf 6}(v_1,v_2)$ and $\chi_{\bf 3}(u)$ of $SU(2)_u\times USp(6)_{\bf v}$ respectively. Moreover, in the index of the $E_7$ SCFT we have two $SU(4)$ symmetries parametrized by ${\bf z}^{(1)}$ and ${\bf z}^{(2)}$ manifestly visible, and the $USp(6)_{\bf v}\times SU(2)_u$ is imbedded in these as,
\be
&&z^{(1)}_1=\frac{\sqrt{u} \sqrt{{v_1}} \sqrt{{v_2}}}{\sqrt{{v_3}}}\,,\;\; z^{(1)}_2=\frac{\sqrt{u} \sqrt{{v_1}} \sqrt{{v_3}}}{\sqrt{{v_2}}}\,,\;\; z^{(1)}_3=\frac{\sqrt{u} \sqrt{{v_2}} \sqrt{{v_3}}}{\sqrt{{v_1}}}\,,\\
&&z^{(2)}_1=\frac{\sqrt{u} \sqrt{{v_3}}}{\sqrt{{v_1}} \sqrt{{v_2}}}\,,\;\; z^{(2)}_2=\frac{\sqrt{u} \sqrt{{v_2}}}{ \sqrt{{v_1}} \sqrt{{v_3}}}\,,\;\; z^{(2)}_3=\frac{\sqrt{u} \sqrt{{v_1}}}{\sqrt{{v_2}} \sqrt{{v_3}}}\,.\nonumber
\ee
Therefore, we conclude that it is possible that the two theories can be linked by going only on this $1d$ subspace. If this is true then the anomalies involving the preserved $SU(2)\times USp(6)$ global symmetry must also match. Indeed it is possible to show that they do. On the $\mathcal{N}=1$ $USp(4)$ gauge theory side, we have as our basic fields five chirals in the $\bold{3}_{SU(2)}$ with $U(1)_t$ charge $-1$ and free R-charge and four chirals in the $\bold{2}_{SU(2)} \bold{6}_{USp(6)}$ with $U(1)_t$ charge $\frac12$ and free R-charge. On the $E_7$ side we have as our basic fields two chirals in the $\bold{4}_{SU(2)}$ with $U(1)$ charge $-1$ and free R-charge and four chirals in the $\bold{2}_{SU(2)} \bold{6}_{USp(6)}$ with $U(1)$ charge $\frac12$ and free R-charge, where we have used the duality of \cite{Argyres:2007ws} to represent the anomalies of the rank $1$ $E_7$ SCFT in terms of free chiral fields. It is straightforward to show that indeed all the anomalies match.

\

\section{Dual of $E_8$ MN theory with $USp(4)$ subgroup gauged}

We consider yet another example where one side is an $\mathcal{N}=1$ conformal gauge theory with a weak coupling limit, while the other contains an intrinsically strongly interacting part, which here is the rank $1$ $E_8$ MN theory \cite{Minahan:1996cj}. The gauge theory side has gauge group $Spin(7)$, ten chiral fields in the spinor representation and five chiral fields in the fundamental representation. With this matter content the gauge one loop beta function vanishes. The theory has a non-anomalous global symmetry of $U(1)_t\times SU(5)\times SU(10)$, and also has classically marginal operators made from a contraction of the vector and two spinor chirals. We assign $U(1)_t$ charge $+\frac12$ to the spinors and charge $-1$ to the vectors. This marginal operator, $\lambda_i^{\alpha\beta}$, is in the $(\bold{5}, \bold{45})$ of $SU(5)\times SU(10)$  (with $i$ being the $SU(5)$ index and $\alpha$ and $\beta$ the $SU(10)$ indices). It is possibe to show that this operator has a non-trivial K\"ahler quotient, and so this theory exists as an SCFT with a conformal manifold containing the weak coupling point. 
We can decompose $SU(10)$ into $SU(2)^5\times U(1)^4$ such that,
\be
{\bf 10} =\sum_{i=1}^5{\bf 2}_i \times a_i\,,\qquad\qquad \prod_{i=1}^5 a_i=1\,, 
\ee where $a_1,\cdot, a_4$ are the fugacities of the $U(1)^4$. Identifying the $a_i$ with the (one over the square root of) Cartan of $SU(5)$ and turning on only the operators which are invariant under this $SU(2)^5 \times U(1)^4$ we get a one dimensional sub-locus of the conformal manifold. To see that we decompose all the marginal operators and conserved currents into representations of $SU(2)^5\times U(1)^4$ to obtain,
\be
&&(\bold{5}, \bold{45})\to \sum_{l=1}^5 a_l^{-2} \left(\sum_{i< j}{\bf 2}_i\times {\bf 2}_j \times a_i a_j+\sum_{i=1}^5 a_i^2\right)\,,\\
&&{\bf 24}+{\bf 99} \to 4+\sum_{i\neq j} a^2_i/a^2_j+\sum_{i=1}^5 {\bf 3}_i+4+\sum_{i\neq j} {\bf 2}_i\times {\bf 2}_j \times a_i/a_j\,.\nonumber
\ee Subtracting the conserved currents from the marginal operators we obtain,
\be
1+\sum_{i< j}\sum_{l\neq i,j}  {\bf 2}_i\times {\bf 2}_j \times \frac{a_i a_j}{a_l^2}-4-\sum_{i=1}^5 {\bf 3}_i\,,
\ee which corresponds to one exactly marginal operator preserving $SU(2)^5\times U(1)^4$ and a collection of marginal operators which break these symmetries.
We can continue to break the symmetry gradually. Turning on any one of the charged marginal operators will break the two involved $SU(2)$ groups to the diagonal and will break one combination of the $U(1)$ symmetries. Turning on the five operators ${\bf 2}_i\times {\bf 2}_{i+1}\times \frac{a_ia_{i+1}}{a_{i+2}^2}$ (where we identify indices $\text{mod}\, 5$) we break all the $SU(2)$ symmetries to the diagonal  and break all the $U(1)$ symmetries (except for $U(1)_t$).  We are also left with many adjoint operators of the diagonal $SU(2)$, turning one of which we break the $SU(2)$ to the Cartan, and then turning additional operators charged under the Cartan we can break the symmetry completely. In the end
 the $SU(5)\times SU(10)$ has been completely broken on the conformal manifold leading to a $5 \times 45 - 99 - 24 = 102$ dimensional conformal manifold, on a generic point of which only a $U(1)_t$ is preserved. The theory has $70$ chiral operators of dimension $2$ given by the symmetric invariant of the vector chiral fields, transforming in the $\bold{15}$ of $SU(5)$ and with charge $-2$ under $U(1)_t$, and the symmetric invariant of the spinor chiral fields, transforming in the $\bold{55}$ of $SU(10)$ and with charge $+1$ under $U(1)_t$. 

The dual side is an $\mathcal{N}=1$ $USp(4)$ gauging of the $\mathcal{N}=2$ rank one SCFT with $E_8$ global symmetry with six chiral fields in the fundamental representation for the $USp(4)$. As the $E_8$ SCFT provides an effective number of twelve fundamental chirals for the $USp(4)$ beta function, the latter vanishes. The theory has a $U(1)_t \times SU(6)\times SO(11)$ global symmetry. Here the $SU(6)$ is the symmetry rotating the $USp(4)$ doublets and $SO(11)$ is the commutant of $USp(4)$ inside $E_8$. The $U(1)_t$ group is the anomaly free combination of the $U(1)$ acting on the four $USp(4)$ doublets and $U(1)_t$, which is the commutant of the $\mathcal{N}=1$ $U(1)_R$ in the $\mathcal{N}=2$ $U(1)_R \times SU(2)_R$. It is straightforward to show that under $U(1)_t$ the $USp(4)$ doublets have charge $-1$ where we have normalized $U(1)_t$, as before, such that the moment map operators of the $E_8$ SCFT have charge $+1$.

We have relevant operators of dimension two given by the moment maps of the $E_8$ SCFT which transform in the $\bold{248}_{E_8}$. After the gauging these decompose to $USp(4) \times SO(11)$ according to $$\bold{248}_{E_8} \rightarrow \bold{10}_{USp(4)} + \bold{5}_{USp(4)}\bold{11}_{SO(11)} + \bold{55}_{SO(11)} + \bold{4}_{USp(4)}\bold{32}_{SO(11)}.$$ In particular, we have the gauge variant $\bold{4}_{USp(4)}\bold{32}_{SO(11)}$ operators, which can be made into a dimension three gauge invariant operators  via a contraction with the $USp(4)$ chiral doublets. This gives a classically marginal operator $\lambda_i$ (with the $i$ being the $SU(6)$ index) in the $\bold{6}_{SU(6)}\bold{32}_{SO(11)}$. Additionally, as the moment map operators carry charge $+1$ under the non-anomalous $U(1)_t$ and the chiral $USp(4)$ fundamentals carry charge $-1$, it is uncharged under the $U(1)_t$. We note that as ${\bf 32}$ contains a singlet in its sixth completely antisymmetric  power, taking $\e\cdot \lambda^6$ contains a singlet meaning the K\"ahler quotient is not empty. There are different possible choices of symmetries to preserve, leading to many different subspaces. One choice is to break the $SU(6)\times SO(11)$ symmetry to $U(1)^2 \times SU(2) \times SU(3)$, where we break $SU(6)\rightarrow U(1)\times SU(4)\times SU(2) \rightarrow U(1)^2\times SO(4) \rightarrow SU(2) \times U(1)^3$ and $SO(11)\rightarrow U(1)\times SU(5)\rightarrow U(1)^2 \times SU(2) \times SU(3)$. The $SU(2)$ is then the diagonal one and the $U(1)^2$ is a combination of the various $U(1)$ commtants. It is possible to show, with methods similar to those previously used, that this leads to a $1d$ subspace preserving $U(1)_t\times U(1)^2 \times SU(2) \times SU(3)$ global symmetry. We can then continue and further break all the $U(1)^2 \times SU(2) \times SU(3)$ part of the global symmetry, leading to a $6 \times 32 - 55 - 35 = 102$ dimensional conformal manifold, on a generic point of which only $U(1)_t$ is preserved.

The theory has $70$ dimension two operators, $55$ of which are given by the moment map operators associated with the $SO(11)$ and are in the $\bold{55}$ of $SO(11)$ and have $U(1)_t$ charge $+1$. The remaining $15$ operators come from the antisymmetric invariant of the $USp(4)$ doublets, transform in the $\bold{15}$ of $SU(6)$ and carry charge $-2$ under the $U(1)_t$. Note that as the $SU(6)$ did not originate from an $\mathcal{N}=2$ theory, it does not have moment map operators.

The $E_8$ SCFT has conformal anomalies, 
 \be
 a=\frac{95}{24}\,,\qquad c=\frac{31}6\,.
 \ee These are the anomalies which can be obtained from $11$  free vectors and $91$ free chiral fields. We add to the model $10$ gauge fields of $USp(4)$ and additional six fundamental fields, the number of which is $24$. The conformal anomalies of the theory are thus, $a=\frac{19}{3}$ and $c=\frac{89}{12}$. If we are after a conformal dual of this model it has to have,
\be
\text{dim}\,{\frak G}=11+10=21\,,\qquad \qquad \text{dim}\,{\frak R}=91+24=115\,.
\ee Having $\text{dim}\,{\frak G}=21$ and assuming the dual is a conformal Lagrangian theory we have several options for a  candidate gauge group: $USp(6)$, $Spin(7)$, $SU(2)^7$, $SU(4)\times SU(2)^2$, $USp(4)\times SU(3)\times SU(2)$. We find such a dual mentioned above with a $Spin(7)$ gauge group.
This model has $5$ vector fields and $10$ spinor fields. This matter content amounts to $10\times 8+7\times 5=115$ free chiral fields, guaranteeing that the conformal anomalies match. 

We can also easily compare the 't Hooft anomalies for the symmetry preserved on a generic point of the conformal manifold. The $Spin(7)$ theory has $80$ free fields with charge $+\frac12$ coming from the spinors and $35$ free fields with charge $-1$ coming from the vectors. To figure out the anomalies of the $E_8$ SCFT we use an Argyres-Wittig duality \cite{Argyres:2007tq}. The $E_8$ SCFT with a $USp(4)$ subgroup of $E_8$ gauged with an ${\cal N}=2$ vector multiplet is dual to a $USp(6)$ ${\cal N}=2$ gauge theory with a half-hypermultiplet in ${\bf 14}$ and eleven half-hypermultiplets in the ${\bf 6}$. This means that the $E_8$ SCFT has the same anomalies as $14+11\times 6=80$ free chiral fields with $U(1)_t$ charge $+\frac12$ and $21-10=11$ free chiral fields with $U(1)_t$ charge $-1$.
We add $24$ more free fields with charge $-1$ (the fundamentals of $USp(4)$) when we gauge the $USp(4)$ to obtain our model. Thus in total we have $80$ free fields with charge $+\frac12$ and $35$ free fields with charge $-1$. We thus get perfect agreement between the two sides of the duality.

So far we have seen that both theories exist as interacting SCFTs with a conformal manifold. We have also seen that the dimension of the conformal manifold, generically preserved global symmetry, anomalies for these symmetries, and relevant operators all match between the two theories. This prompts us to propose that these two theories are in fact dual and share the same conformal manifold. The global symmetry at the weak coupling point differs, but this can easily be accounted for as most the global symmetry is broken when moving on the conformal manifold. The $U(1)_t$ group is the only part that is never broken and so must match between the two theories. We next compare the indices of the two theories presenting additional evidence for our claim.

Here, the full index of the $E_8$ SCFT is not yet determined.
However one can compute the Schur limit of the index. We  use \cite{Gadde:2011ik,Gadde:2011uv} to write the Schur index of the $E_8$ SCFT as\footnote{Here, as in the $E_7$ case, as $U(1)_t$ is preserved we can compute other ${\cal N}=2$ limits of the index, and in particular the Coulomb limit. On the $E_8$ side we have a contribution from the dimension six Coulomb branch operator $1/(1-x^6)$ and a contribution from the gauging $\oint\frac{dz_1}{2\pi i z_1} \oint\frac{dz_2}{2\pi i z_2} \Delta_{USp(4)}(z_1,z_2)\frac1{((1-x z_1^{\pm1})(1-x z_2^{\pm1}))^6}=\frac{1-x^6}{(1-x^2)^{15}}$. While on the $Spin(7)$ side the only fields which contribute are the vectors, and we get 
$\oint\frac{dz_1}{2\pi i z_1} \oint\frac{dz_2}{2\pi i z_2} \oint\frac{dz_3}{2\pi i z_3}\Delta_{Spin(7)}(z_1,z_2,z_3)\frac1{((1-x)(1-x z_1^{\pm2})(1-z_2^{\pm1})(1-z_3^{\pm1}))^5}=\frac1{(1-x^2)^{15}}$. The two indices manifestly match. },

\be
\hskip-1.in&&I_{E_8}(z_1,z_2) =\\
&&\frac{(qz_1^{\pm1}z_2^{\pm1};q)^{-2}\prod_{i=1}^2(q z_i^{\pm1};q)^{-4}(qz_i^{\pm2};q)^{-1}}{(1-q)^5(1-q^2)^4(1-q^3)^3(1-q^4)^2(1-q^5)(q^3;q)^{4}(q^2;q)^{13}(q;q)^{13}} \sum_{\lambda_1=0}^\infty \sum_{\lambda_2=0}^{\lambda_1}\sum_{\lambda_3=0}^{\lambda_2}\sum_{\lambda_4=0}^{\lambda_3}\sum_{\lambda_5=0}^{\lambda_4}\nonumber\\
&&
\frac{\chi_{\{\lambda_i\}}(q^{\frac12},q^{\frac12},q^{\frac12},q^{-\frac12},q^{-\frac12},q^{-\frac12})\chi_{\{\lambda_i\}},(1,1,q,q,q^{-1},q^{-1})\chi_{\{\lambda_i\}}(1,1,z_1,z_1^{-1},z_2,z_2^{-1})}{\chi_{\{\lambda_i\}}(q^{\frac52}q^{\frac32},q^{\frac12},q^{-\frac12},q^{-\frac32},q^{-\frac52})}\,.\nonumber
\ee Here we only refine the index with the fugacities for the  $USp(4)$ symmetry we are going to gauge. Note that the construction of the index has manifest $SU(6)\times SU(2)\times SU(3)$ subgroup of $E_8$ and $USp(4)$ is imbedded in the $SU(6)$ such that $SU(6)\to SU(2)\times U(1)\times SU(4)$ followed by $SU(4)\to USp(4)$ with ${\bf 4}_{SU(4)}\to {\bf 4}_{USp(4)}$. As before the $\lambda_i$ are the lengths of the rows of the Young tableaux defining $SU(6)$ representations and $\chi_{\{\lambda_i\}}$ are the corresponding Schur polynomials.
Then we define the single letter partition function of the extra fields on the $E_8$ side of the duality to be,

\be
M_A(z_1,z_2;q)=\frac{q^{\frac12}}{1-q}\left(6\chi_{\bf 4}(z_1,z_2)\right)-\left(\frac{q}{1-q}+\frac{q^{\frac12}}{1-q^{\frac12}}\right)\chi_{\bf 10}(z_1,z_2)\,,
\ee giving the index,

\be
&&I_A = \oint\frac{dz_1}{2\pi i z_1}\oint\frac{dz_2}{2\pi i z_2}  \Delta_{USp(4)}(z_1,z_2)I_{E_8}\left(z_1,z_2\right) PE\left[M_A(z_1,z_2;q)\right]\,.\nonumber
\ee On the quiver side of the duality the contribution of the matter is,

\be
&&M_B(z_1,z_2,z_3;q)=\frac{q^{\frac12}}{1-q}\left(5\chi_{\bf 7}(z_1,z_2,z_3)+10\chi_{\bf 8}(z_1,z_2,z_3)\right)-\left(\frac{q}{1-q}+\frac{q^{\frac12}}{1-q^{\frac12}}\right)\chi_{\bf 21}(z_1,z_2,z_3)\,,\nonumber
\ee with the index given by,

\be
I_B =\oint\frac{dz_1}{2\pi i z_1} \oint\frac{dz_2}{2\pi i z_2} \oint\frac{dz_3}{2\pi i z_3} \Delta_{Spin(7)}(z_1,z_2,z_3)\, PE\left[M_B(z_1,z_2,z_3;q)\right]\,.
\ee Let us write explicitly the characters of represnetations of $Spin(7)$ for completeness,
\be
&&\chi_{{\bf 7}}(z_1,z_2,z_3)=1+\sum_{i=1}^3z_i^{\pm2}\,,\qquad \chi_{{\bf 21}}(z_1,z_2,z_3)=\frac12\left(\chi_{{\bf 7}}(z_1,z_2,z_3)^2-\chi_{{\bf 7}}(z^2_1,z^2_2,z^2_3)\right)\,,\nonumber\\
&&\chi_{{\bf 8}}(z_1,z_2,z_3)=z_1z_2z_3\,\left(1+z_1^{-1}+z_2^{-1}+z_3^{-1}\right)+z_1^{-1}z_2^{-1}z_3^{-1}\left(1+z_1+z_2+z_3\right)\,.
\ee
The result of the computation in the two duality frames is equal and is given by,

\be
&&{\cal I}_A={\cal I}_B= 1+70 q+171 q^{\frac32}+2715 q^2+11405 q^{\frac52}+85725 q^{3}+411873 q^{\frac72}+\\
&&\;\;\;2306124 q^{4}+10863905 q^{\frac92}+52351904 q^{5}+231967709 q^{\frac{11}2}+1012822602 q^{6}+\cdots\,.\nonumber
\ee

We thus have gathered a compelling collection of evidence that the two theories are indeed dual to each other.

\section*{Acknowledgments}

We are grateful to Evyatar Sabag for relevant discussions.
The research of SSR is supported by Israel Science Foundation under grant no. 2289/18, by I-CORE  Program of the Planning and Budgeting Committee, and by BSF grant no. 2018204. GZ is supported in part by the ERC-STG grant 637844-HBQFTNCER and by the INFN.

\appendix

\section{${\cal N}=1$ Schur index}\label{app:schur}

Let us discuss basic definitions of the supersymmetric index and its Schur limit specification. We use the standard definitions of the index \cite{Kinney:2005ej,Romelsberger:2005eg,Dolan:2008qi} which can be found in {\it e.g.} \cite{Rastelli:2016tbz}. Concretely the index is the trace over the Hilbert space of a $4d$ ${\cal N}=1$ theory quantized on ${\mathbb S}^3$ and it depends on two parameters, $q$ and $p$, and on a set of fugacities for global symmetries,

\be
{\text{Tr} } (-1)^F q^{j_2-j_1+\frac12 R} p^{j_2+j_1+\frac12 R} \prod_{i=1}^{\text{dim} \, G_F} a^{q_i}_i\,.
\ee Here $j_i$ are the Cartan generators of the two $SU(2)$ isometries of the sphere and $R$ is the R-charge assignment. The group $G_F$ is the global symmetry group, $a_i$ are fugacities of the Cartan maximal torus of $G_F$, and $q_i$ are the charges under these abelian symmetries. The states contributing to the index satisfy, $E-2j_2-\frac32 R=0$, with $E$ being the scaling dimension.  This combination picks up one of the four supercharges relative to which the index is computed.

The index can be computed by projecting on gauge invariant states the contributions of chiral and vector fields. The contributions of the fields can be written using the plethystic exponent,

\be
\text{PE} [ f(a,b,\cdots)]= \text{exp}\left(\sum_{l=1}^\infty \frac1l f(a^l,b^l,\cdots)\right)\,.
\ee The index of the chiral field of R-charge $R$ and represnetation ${\frak R}$ under  symmetry group is given by,

\be
{\cal I}_R({\bf z})=\text{PE}\left[ \frac{(qp)^{\frac{R}2} \chi_{\frak R}({\bf z})-(pq)^{1-\frac{R}2} \chi_{\overline{\frak R}}({\bf z})}{(1-p)(1-q)}\right]\,
\ee while the contribution of the vector amounts to,

\be
{\cal I}_v({\bf z})=\text{PE}\left[-\left(\frac{q}{1-q}+\frac{p}{1-p}\right)\chi^{G}_{adj}({\bf z})\right]\,,
\ee where $G$ is the gauge group and $\chi^G_{adj}$ is the character of the adjoint representation. Note that the numerator in the plethystic exponent comes from zero modes of fields, bosons with plus sign and fermions with minus sign, while the denominator comes from two derivatives which contribute to the index. The argument of the plethystic exponent is usually called the single letter index.

Note that computing the plethystic exponent for the chiral field the result will be a double infinite product giving rise to an elliptic Gamma function \cite{Dolan:2008qi}. However, if the representation is (pseudo)real, $\chi_{\frak R}=\chi_{\overline{\frak R}}$, then we can choose the parameters $q$ and $p$ to be related as,

\be\label{specSchur}
p = q^{\frac{1-R}{R}}\,,
\ee and then additional cancelation between fermions and bosons occur leading to the index of the chiral to be,

\be
\text{PE}\left[ \frac{q^{\frac12}}{1-q} \chi_{\frak R}({\bf z})\right]\,.
\ee Note that this is precisely the expression for the Schur index \cite{Gadde:2011ik,Gadde:2011uv} of  an ${\cal N}=2$ half-hypermultiplet. In the ${\cal N}=2$ case the extra cancelations in the index can be explained by the index being invariant under additional supercharge. 
For the ${\cal N}=1$ case such cancelation will not in general happen if we have fields of different R-charges and non (pseudo)real representations. However, in the conformal Lagrangian theories such that all the R-charges are free, and in particular are the same, the cancelations in the limit can happen if all the representations are (pseudo)real, as is the case 
in all the examples in this paper. In these cases \eqref{specSchur} becomes $p^2=q$ and we will refer to the limit as ${\cal N}=1$ Schur index. Note that Lagrangian ${\cal N}=2$ theories then are just a special case of this class of theories.

The single letter index of the free vector multiplet becomes,

\be
-\left(\frac{q}{1-q}+\frac{q^{\frac12}}{1-q^{\frac12}}\right)\chi^{G}_{adj}({\bf z})=
-\left(\frac{2q}{1-q}+\frac{q^{\frac12}}{1-q}\right)\chi^{G}_{adj}({\bf z})\,,
\ee and it formally looks as that of an ${\cal N}=2$ vector multiplet (the first term) and the inverse contribution of an adjoint chiral field. If the theory happens to be ${\cal N}=2$ then it will also contain a chiral field in adjoint representation of the gauge group, the contribution of which will cancel with the latter term. If one is to compute the ${\cal N}=1$ Schur index of  a conformal theory constructed by an ${\cal N}=1$ conformal gauging of an ${\cal N}=2$ component with possibly ${\cal N} =1$ additional matter, as is done in this paper, one can first compute the Schur indices of the components and then combine them together. This is useful as, though for most ${\cal N}=2$ theories a Lagrangian is not known yet, we do know for large classes of them the value of the Schur index \cite{Gadde:2011ik,Gadde:2011uv}, and this class of theories was enlarged recently by the discovery of the relation between Schur indices and chiral algebras \cite{Beem:2013sza}. As with the ${\cal N}=2$ Schur index also the ${\cal N}=1$ version can be expressed using $\theta$ functions,

\be
\theta(z;q) =\prod_{l=0}^\infty (1- z q^l)(1-z^{-1} q^{l+1})=(z;q)(z^{-1}q;q)\,,\;\;\; \left(\;\; (z;q)=\prod_{l=0}^\infty (1-z\,q^l)\;\; \right)\,.
\ee Using the fact that the representations we allow are (pseudo)real, the non-zero weights come in $\pm$ pairs. We thus can split the non zero  weights arbitrarily into a group of ``positive'' and ``negative'' roots, ${\frak W}_+$ and ${\frak W}_-$, such that if $w\in {\frak W}^{\frak R}_+$ then $-w\in {\frak W}^{\frak R}_-$. We denote the number of zero weights by $n^0_{\frak R}$. Then the index of the chiral field is,

\be
\frac1{(q^{\frac12};q)^{n^0_{\frak R}}} \prod_{w\in {\frak W}^{\frak R}_+} \frac1{\theta(q^{\frac12} e^w;q)} =
\frac1{(q^{\frac12};q)^{n^0_{\frak R}}} \prod_{w\in {\frak W}^{\frak R}_-} \frac1{\theta(q^{\frac12} e^w;q)}\,.
\ee The index of a gauge theory with group $G$ can be written then as,

\be
\frac1{|W|}(q;q)^{2\text{rank}\, G} (q^{\frac12};q)^{\text{rank}\, G-n^0_{\frak R}} \prod_{i=1}^{\text{rank}\, G}\oint \frac{dz_i}{2\pi i z_i} \prod_{v\in V^G_+} \theta(e^{\pm v};q) \theta(q^{\frac12} e^v;q)\prod_{w\in{\frak W}^{\frak R}_+} \frac1{\theta(q^{\frac12} e^w;q)}\,.  \nonumber\\
\ee Here $W$ is the Weyl group of $G$, the $z_i=e^{e_i}$ parametrize the maximal torus with $e_i$ spanning the space of roots, and $V_+$ is the set of positive roots.
As the index can be written in terms of objects with interesting $SL(2;\Z)$ properties, see {\it e.g.} \cite{Razamat:2012uv,Beem:2017ooy,Cordova:2016uwk}, it is tempting to speculate that also the ${\cal N}=1$ Schur index has to do something with $2d$ CFTs/chiral algebras.

\bibliographystyle{ytphys}
\bibliography{refs}

\end{document}